# A STEP TOWARDS INHERENTLY INTERPRETABLE CAUSAL MACHINE LEARNING MODELS FOR DECISION SUPPORT

*Completed Research Paper*

David Zapata Gonzalez, Paderborn University, Paderborn, Germany, david.zapata@upb.de

## Abstract

*"The growing reliance on machine learning for decisions across sectors underscores the importance of model transparency and interpretability. Existing post-hoc explainability methods and inherently interpretable approaches shed light on model behavior, yet they primarily reveal how models exploit correlations to maximize performance in prediction tasks. However, many decisions require causal insights and the possibility of using models for what-if scenario evaluation. To address this, we propose the integration of causal machine learning with inherently interpretable models for cross-sectional data. We evaluate these methods in terms of predictive accuracy and interpretability. Our findings show that the proposed approach achieves competitive performance in prediction and what-if analysis while offering transparency on the system structure, causal relationships among variables, and the functional forms that connect them. This work contributes to research on causality, machine learning interpretability, and data-driven decision support by offering informed, transparent, and causally grounded decisions."*

## 1 Introduction

Machine learning (ML) is becoming an increasingly integral part of decision-making in organizations, with its role expanding steadily across industries (Abdel-Karim et al., 2021; Jayatilake & Ganegoda, 2021; Jhaveri et al., 2022). As reliance on these systems grows, decision-makers require a deeper understanding of model recommendations and the ability to draw causal insights from them (Barredo Arrieta et al., 2020). In this regard, interpretability and causality are critical for making transparent, reliable, and informed decisions based on ML systems.

Interpretability means that an ML model can explain its decisions in a way people can understand (Doshi-Velez & Kim, 2017). There are two paradigms in interpretable ML: training complex black-box ML models and then using post-hoc explainability methods from the field of explainable artificial intelligence (XAI) (Barredo Arrieta et al., 2020), and the often overlooked approaches that rely on inherently interpretable ML (IIML) (Rudin, 2019; Zschech et al., 2025). Notably, IIML models are directly understandable and can therefore communicate their reasoning more transparently and reliably without needing additional explanation methods (Rudin & Radin, 2019).

Orthogonal to the problem of ML interpretability, regardless of whether post-hoc methods or inherently interpretable models are used, their interpretations only reveal how ML models exploit correlations in data to maximize prediction performance, which should not be mistaken with causal insights (Molnar et al., 2022; Zschech et al., 2025). Furthermore, ML approaches can make predictions only for future observations that follow the same underlying distribution seen during training and may struggle to evaluate what-if scenarios, which is important for business decision-making (Hünermund et al., 2022).

To obtain causal insights from ML models and use them for tasks beyond prediction, it is necessary to adopt principles from causal inference (Feuerriegel et al., 2024; Pearl, 2009; Schölkopf, 2022). In this context, complex ML models can be employed to capture causal relationships forming Causal Machine Learning (CML) (Kaddour et al., 2022). Similarl to the case of non-causal ML, the complex CML models also require interpretation.

In this context, to overcome the limitations of correlational ML in capturing causal relationships and enabling not only prediction but also what-if analysis, and to enhance the interpretability of the models used, we combine concepts of CML and IIML. Based on this integration, we propose the following research questions:

**RQ1:** How can causal inference be combined with inherently interpretable machine learning models to support decision-making?

**RQ2:** How do the models perform in prediction tasks and what-if analysis?

To address **RQ1**, we propose a framework that combines Structural Causal Models (SCM) with IIML methods, namely Generalized Additive Models (GAMs) and Symbolic Regression (SR), for cross-sectional data. To address **RQ2**, we evaluate the performance of these models against non-causal black-box models for prediction tasks and what-if scenarios analysis.

We found that combining these two fields yields models with competitive predictive performance while also enabling robust evaluation of what-if scenarios. Moreover, the integration ensures a high level of transparency by clearly exposing the system's structure, the causal relationships among variables, the underlying modelling assumptions, and the interpretation of the functional relationships that connect them. This clarity ultimately enables more informed and reliable decision support.

These findings contribute to the growing literature on CML by incorporating inherently interpretable models that clarify not only the causal relationships between variables but also their functional forms. They also advance the field of interpretable machine learning by integrating causal knowledge into the selection of variables and the structure of their relationships. Finally, this work contributes to the decision support literature based on ML by enabling the use of models for what-if analysis and promoting high transparency in decision-making processes.

The paper is structured as follows: Section 2 reviews background and related work on CML, interpretability in machine learning, and the principles of GAMs and SR; Section 3 presents the methodology for combining SCMs and IIML models for cross-sectional data; Section 4 shows the experiments, including two evaluations, one using a generated dataset as demonstration of model´s interpretability and the other for benchmarking the approach with external datasets; Section 5 discusses the empirical findings, outlines the limitations of the study and suggests directions for future research, and Section 6 summarizes the main conclusions and contributions.

# 2 Background

## 2.1 Causal machine learning

Supervised ML consists of developing a statistical model that predicts an output based on one or more input variables. The model is trained on historical data and subsequently applied to forecast outcomes for new, unseen instances drawn from the same underlying data distribution. In this setting, the primary objective is accurate prediction, and the model is optimized to minimize the error between its predictions and the true observed values during training (Hastie et al., 2009).

According to Pearl (2009), the use of ML for prediction typically falls within the first rung of the Ladder of causation, called association, or seeing. At this level, models rely on correlations and conditional probabilities derived from observational data to make predictions (e.g., "What is the projected sales volume next quarter given current market trends?"). The second level, intervention, or doing, involves answering "what-if" questions using tools such as do-calculus, causal graphs, or randomized experiments to estimate the effects of deliberate changes in the system. For instance, "What would happen to sales if we increased our marketing investment? The third and highest level, counterfactuals, or imagining, enables reasoning about hypothetical scenarios that did not occur, for example, "Would sales still have gone up without the investment in marketing?" (Pearl, 2018).

Causal relationships among variables can be represented using Directed Acyclic Graphs in the form of Graphical Causal Models (Pearl, 2010). In a causal graph, a variable that has a directed arrow to another is called a *parent*, while a variable that receives an arrow is called a *child.* The parents are the causes of the childs. A variable can be both a parent to some variables and a child of others. The relationships in a graph offer a structured way to encode assumptions about the data-generating process, capturing both direct and indirect dependencies between variables.

In this regard, the causal graph has basic structures known as chains, forks, and colliders, shown in Figure 1. In the example of a chain, conditioning on Y blocks the relationships between X and Z; in a fork, adjusting for the common cause X removes spurious association between Y and Z; and in a collider, conditioning on the shared effect Z introduces bias by creating a spurious correlation between otherwise independent variables X and Y (Molak, 2023).

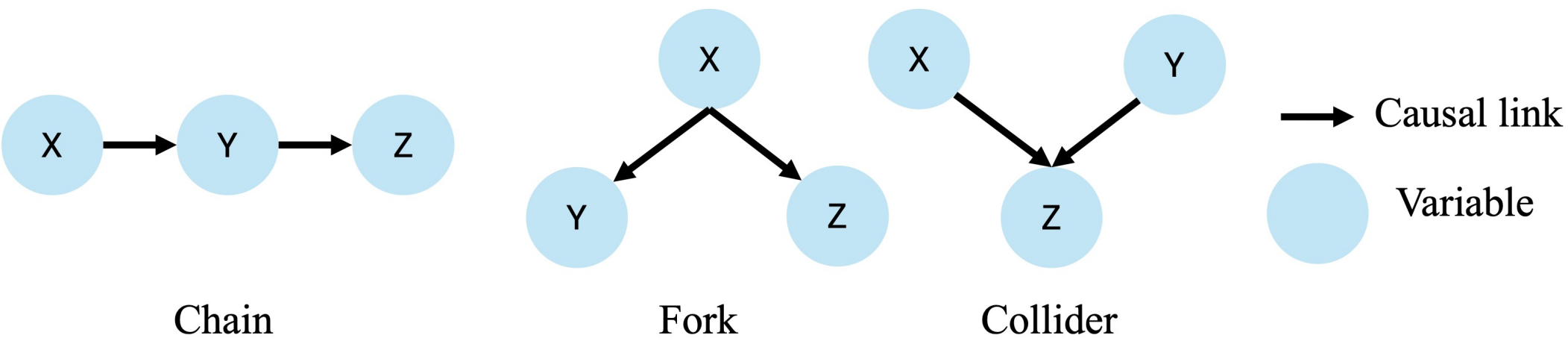

*Figure 1.* *Basic structures in causal graphs.*

Variables can interact in complex networks that involve many interconnected nodes, but the fundamental principles underlying chains, forks, and colliders remain consistent. Conditional independence, through the idea of blocking paths, serves as a key criterion for determining when it is appropriate to control for certain variables in order to reduce biases (e.g., confounding or collider bias), avoid spurious associations, and reveal genuine causal relationships (Peters et al., 2017).

Building on this foundation, once the relevant relationships between variables have been established in a causal graph, it becomes possible to model the functional dependencies among them. This can be done using approaches such as additive linear models or complex ML methods capable of capturing nonlinear effects and interactions (Pearl, 2009, 2014). This process results in a Structural Causal Model (SCM), which represents the causal relationships among variables and their functional dependencies.

The SCMs enable a wide range of causal inference tasks, such as estimating treatment effects, identifying root causes, and predicting what-if scenarios. (Blöbaum et al., 2022; Molak, 2023; Sharma & Kiciman, 2020). Later, post-hoc methods can be used to interpret these complex ML models, including variations of Shapely Additive exPlanations (SHAP) (Carloni et al., 2025; Heskes et al., 2020).

## 2.2 Interpretable machine learning

Interpreting the recommendations of ML systems has become essential for effective decision support, ensuring that models align with domain knowledge, are fair and ethical, safe, and trustworthy (Burkart & Huber, 2021; Meske et al., 2022; Rudin & Radin, 2019). This is especially important in situations where model errors can lead to severe consequences, such as in high-stakes decisions, healthcare, legal settings, or control systems (Bharati et al., 2024; Machlev et al., 2022; Rudin, 2019).

To tackle the problem of model comprehensibility, there are two main approaches to understanding the recommendations of ML systems: explainability methods, often referred to as Explainable AI (XAI), and the use of inherent interpretable ML models (IIML) (Barredo Arrieta et al., 2020). In the context of explainability, the model operates as a black box; its internal mechanisms are too complex to be directly understood by humans. Consequently, additional post-hoc techniques are applied to interpret its behavior with the use of surrogate, simpler models or methods that quantify how individual features contribute to the final prediction, such as SHAP (S. M. Lundberg & Lee, 2017) and Local Interpretable Model-agnostic Explanations (LIME) (Ribeiro et al., 2016). In contrast, inherently interpretable models are designed so that their internal logic is transparent and directly accessible to humans (Rudin, 2019).

In this regard, IIML models provide transparency by design, ensuring that their recommendations are directly understandable and faithfully reflect how predictions are generated, without the need for additional tools to make their outputs comprehensible to humans (Rudin, 2019; Stiglic et al., 2020).

There is a prevailing preference for post-hoc explainability, driven by the assumption that an inevitable trade-off exists between model performance and interpretability. However, this trade-off is not necessarily present in the context of tabular data, unlike in unstructured data such as images or natural language, and IIML can achieve competitive performance compared to their black-box counterparts (Rudin, 2019; Rudin & Radin, 2019).

Nonetheless, because ML models are designed to map input values to outputs by minimizing predictive error (Hastie et al., 2009), the interpretability of a model's decisions, regardless of how transparent they might be, does not imply causal understanding, and for that is necessary to use a causal framework (Zschech et al., 2025, Lundberg et al., 2021). A causal framework, such as the SCM explained in the previous subsection, aims to reduce common sources of bias in data-driven methods and prevent the models from bias and exploiting spurious correlations in data. They also make it possible to move beyond prediction tasks toward causal analysis, such as evaluating what-if scenarios (Pearl, 2018).

## 2.3 Inherently interpretable machine learning models

The simplest inherently interpretable model is a linear regression, where the dependent variable is expressed as an additive combination of the independent variables, assuming a linear relationship between the independent variables and the dependent variable (e.g. $f(x) = aX_1 + bX_2 + cX_3 + \cdots + X_n + e$). Due to its simplicity, these models might be unable to capture intricate relationships between predictors and targets. However, other inherently interpretable models are capable of handling more complex data; in this work, we focus on Generalized Additive Models (GAMs) and Symbolic Regression (SR) due to their success in several fields like health (Bohlen et al., 2025), environment (Pedersen et al., 2019), and scientific discovery (Makke & Chawla, 2024).

In contrast to a linear regression, a GAM replaces the strict linear form with smooth functions (e.g. $f(x) = f_1(X_1) + f_2(X_2) + f_3(X_3) + \cdots + f_n(X_n) + e$) and uses fitting techniques such as penalized regression splines, while controlling complexity through smoothing parameters (Hastie & Tibshirani, 1986). Consequently, although the individual functions might be complex, the additive structure of the model allows to visualize directly each variable´s shape and contribution to the predicted outcome directly, making them inherently interpretable.

Another approach is Symbolic Regression (SR), a machine learning technique that discovers mathematical expressions directly from data (Schmidt & Lipson, 2009). It does so by constructing candidate equations from basic mathematical operators (e.g.,$+$, $-$ $*$,$\div$, $sin$, $log$, $exp$, $ln$) and evaluating their performance using a fitness score like prediction error (Makke & Chawla, 2024).

Because the search space for the right expressions grows rapidly with the number of variables and operators, exhaustively exploring all possible equations is computationally infeasible (Virgolin & Pissis, 2022). Therefore, heuristic search methods are used, such as genetic programming, which allow for efficient exploration of promising regions of the search space and combine high-performing functions candidates. The ultimate goal is to identify the simplest and most accurate equation that explains the data while remaining computationally tractable (Angelis et al., 2023; Dong & Zhong, 2025).

Figure 2 shows a data-generating process (a) for a dependent variable "Y" with two predictors "X1" and "X2". The equation in (b) shows the result of fitting a linear regression to the data, (c) depicts a GAM, and (d) a SR. The partial dependence plots show that for X1, all models achieved a proper fit, but for X2, the GAM and SR were the ones able to capture the relationships well. It can also be seen that the SR shows a symbolic expression, which in this case closely matches the data-generating process.

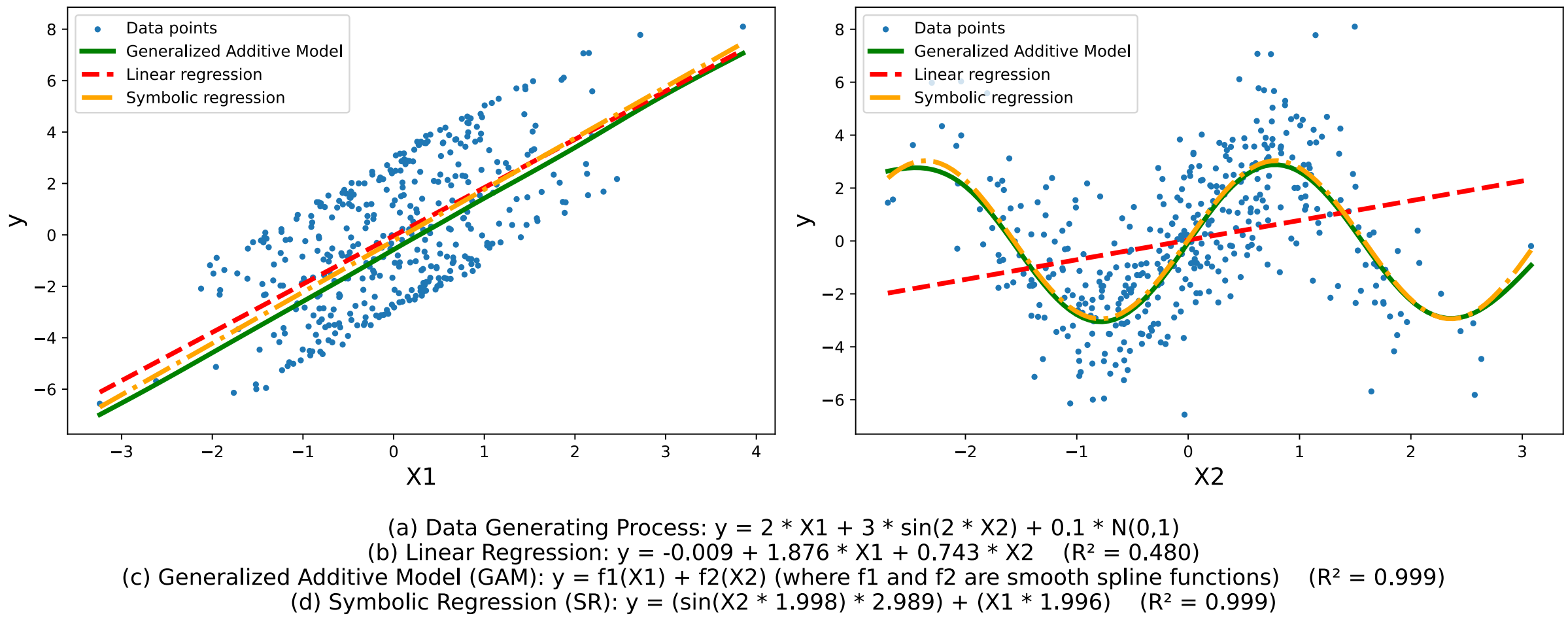

*Figure 2. Partial dependence plots and equations for inherently interpretable models.*

Although GAMs and SR models offer advantages in interpretability, they also have several limitations. Their outputs can be difficult to interpret when interactions exist between variables, or in the case of SR, when the mathematical expressions become highly nested and complex. In addition, these models can be computationally expensive to train (Reinbold et al., 2021; Zschech et al., 2025).

# 3 Methodology

Building on these foundations, we introduce a framework that combines SCMs with IIML models to achieve what-if analysis and inherent interpretability. The framework is shown in Figure 3 and is organized into four parts. The first is the input information, which includes the dataset and a causal graph describing the causal relationships. This causal graph can be constructed using domain knowledge, causal discovery methods, or a combination of both. The second component is the modelling approach, where the black box symbolizes a complex ML model and the white box with the "*f*" symbol denotes an IIML model. There are four modelling options: (a) and (b) correspond to approaches that use all variables to predict the target, either with a black-box or an IIML model, respectively; the third approach (c) shows modelling the causal graph with black-box ML models; and the last option (d) does the same but with IIML models.

The third part of the framework concerns the type of task that each modelling approach supports. Purely data-driven approaches enable only predictive tasks, whereas integrating the causal graphs allows for both prediction and what-if analysis. Finally, the fourth part shows the possible interpretation, which ranges from association post-hoc interpretation of black-box models to a causal inherent interpretation. Overall, we argue that progressing from the lower to the higher levels leads to improved decision support and a deeper understanding and transparency of the ML models.

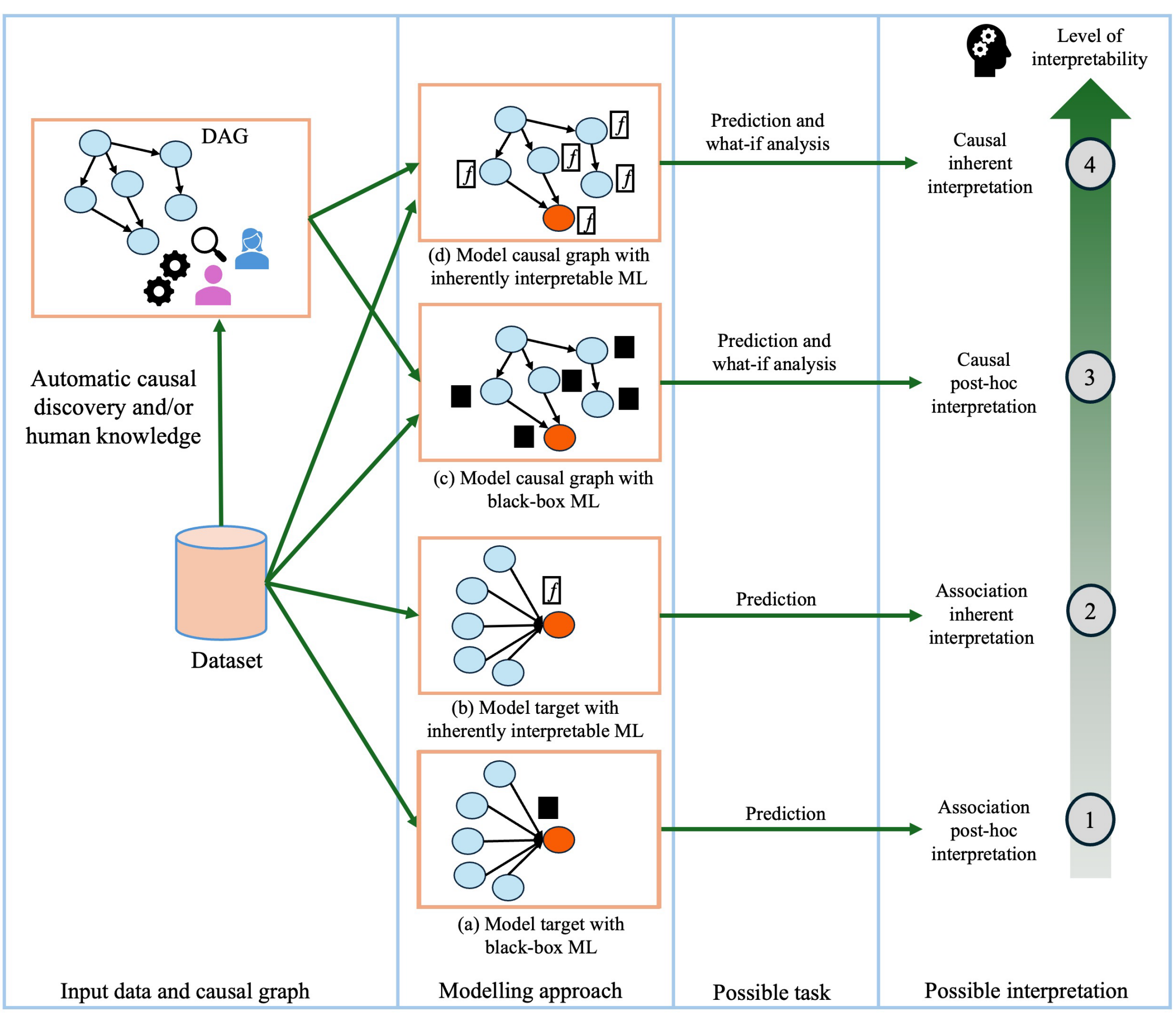

*Figure 3. Proposed framework. A combination of causal graphs with IIML models.*

The following four subsections provide a more detailed explanation of each part of the framework, and the fifth describes the evaluation procedure.

## 3.1 Input data and causal graph

As with any ML prediction task, we begin by preprocessing the dataset according to the chosen modelling approach. In the context of CML, however, there is an additional step: constructing a causal graph that represents the causal relationships among variables. These graphs serve as the backbone of causal reasoning by providing structure for causal identifiability and transparency about modelling assumptions.

A causal graph can be constructed based on domain expertise. Such a graph encodes expert knowledge about the system while making key assumptions explicit, for example, the absence of strong unobserved confounders and the acyclicity of the relationships among variables. Also, causal discovery algorithms can be used to infer causal graphs directly from observational data (Chickering, 2020; Glymour et al., 2019; Shimizu, 2014; Zheng et al., 2018). However, because these methods depend on specific assumptions, the resulting structures may not always be accurate (Hasan et al., 2024). It is therefore important to complement algorithmic outputs with domain knowledge to assess and improve their quality.

In addition, several approaches have been proposed to evaluate the reliability of inferred causal graphs (Eulig et al., 2025; Faller et al., 2024). Moreover, the causal graph can still be used even when certain assumptions are violated and there are also methods specifically designed to address such violations, and mitigate their impact (Cinelli & Hazlett, 2020).

In summary, there are tools available to support the extraction and verification of causal graphs, but ultimately, it is not possible to confirm the exact correctness of a causal graph in practical applications. However, we argue that this uncertainty should not discourage its use. The graph represents the best available understanding of the world based on human knowledge, makes the underlying assumptions about the system explicit, and enables more transparent ML modelling and informed decisions—decisions that otherwise rely solely on the implicit assumptions of ML developers and decision-makers.

## 3.2 Modelling approach

In CML, once a causal graph has been identified, it becomes possible to model the functional relationships for the variables in the graph. The modelling can be implemented using black-box algorithms or IIML models.

In the SCM framework, each variable $V$ is modelled as a function of its parents, $\mathrm{PA}(V)$ and an exogeneous noise term $(U_i)$, $V_i = f_i(\mathrm{PA_i}, U_i))$ (Pearl, 2009). Root nodes are treated as exogenous variables, modelled solely through their noise term rather than fitted predictive models of other variables.

An example of this setup is shown in Figure 4 (a), which depicts an SCM with 5 variables, causal relationships, and structural equations. When modeling a variable, we include only its direct parents as inputs. In the case of Y, only Z and B would be used as input variables. In turn, Z would require its parents, X and A. In this case, including X when modelling Y is unnecessary since, according to the causal structure, the effect of X on Y operates entirely through Z, and once Z is controlled for, Y becomes conditionally independent of X.

In contrast, an ML approach typically employs all available or automatic feature selection based on correlations and predictive power, regardless of causality. Here, we would use a single model to predict the target Y. Although not typically represented as a graph, it can be conceptually shown in Figure 4 (b).

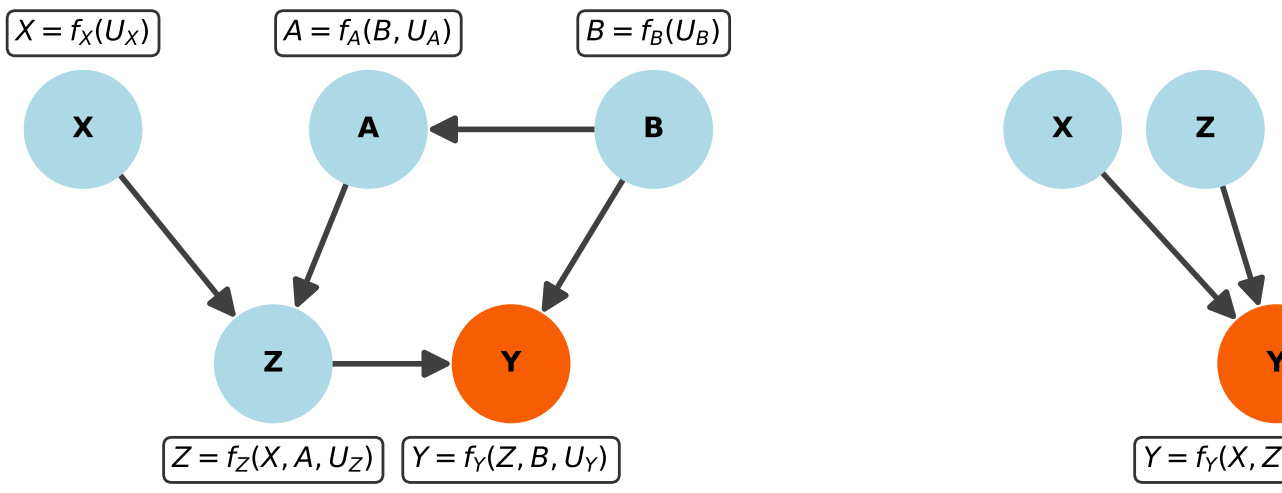

(a) Causal graph with structural equations. (b) Modelling with all variables.

*Figure 4. Two modelling approaches for cross-sectional data.*

When we use the graph structure to choose variables, we often need to train more models, one for each node that has parents. However, each of these models relies on fewer variables than a standard ML approach, which can make them individually simpler and easier to interpret.

Moreover, the causal approach tries to emulate the data-generating process and enables not only prediction tasks but also "what-if" analysis. For instance, if we wish to evaluate what would happen to Y under an intervention that changes variable A while keeping the others fixed, we would block all incoming paths to A (e.g., A → B) and then propagate the predictions through the causal graph. In this case, we would first use X and the new (intervened) value of A to predict Z, and then use the results of Z together with B to predict Y.

In contrast, under the ML approach illustrated in Figure 4 (b), changing the value of A while keeping the other fixed would lead to an invalid "what-if" analysis. This is because Z is causally influenced by A, meaning that an intervention on A would also affect Z; the values of this variable have also changed under the intervention, and this must be incorporated into the modelling of Y. Ignoring this causal dependency violates the system's underlying causal structure, leading to erroneous results.

## 3.3 Possible Task

In purely predictive tasks, incorporating all available features into a machine learning model, even those that are only spuriously correlated with the target, can improve predictive performance, making such models highly effective for prediction. However, such models are less robust under distributional shifts or out-of-distribution predictions and, as explained in the previous subsection, are not useful for what-if scenarios (Schölkopf, 2022).

In the case of CML, what-if analysis differs from prediction tasks. For what-if analysis, we need to change the intervening variables and propagate the effect across the entire graph; the prediction of one model is the input of another model until the target node is achieved. However, for pure prediction tasks, when all predictors are observed and only the target variable is missing, it is not necessary to propagate the predictions through the entire causal graph, since this leads to the cumulation of prediction errors. Instead, we perform a causal feature selection (Yu et al., 2021), making predictions only on the target variable based on observed values of its causal parents.

## 3.4 Possible Interpretation

For the interpretation of the models, one can either use post-hoc methods or directly analyze IIML models. When working with the ML approach, it is crucial to distinguish between the associations they learn and the interpretations derived from them, since these models select and combine features primarily to maximize predictive performance, and their learned relationships do not necessarily reflect causal relationships. In contrast, combining causal graphs with IIML models provides maximum transparency by (1) offering a clear understanding of the causal variables involved, (2) an explicit representation of the structural relationships among them, and (3) an interpretable functional form that describes how these variables influence each other.

## 3.5 Evaluation

For the evaluation[1] of the framework, we compare models trained on all available variables **(ML)** with those trained according to the graph structure **(CML)**. As black-box models, we use XGBoost regressor **(XGBoost)** from the XGBoost library (Chen & Guestrin, 2016), Random Forest **(RForest)** and Multilayer Perceptron (**MLP**) from the Scikit-learn library (Pedregosa et al., 2011). For inherently interpretable models, we include Linear Regression **(LinReg)** from Scikit-learn, Generalized Additive Models **(GAM)** from Servén & Brummitt, (2018), as well as Symbolic Regression **(SR)** from Cranmer (2023). We perform hyperparameter tuning for all the models using extensive random search and cross-validation ($K = 4$). The GAMs are trained without interaction between features and using smooth splines, and the SR can use the following mathematical operators: "sin", "log", "exp", "ln", "+", "*", "-", "÷".

Finally, we conduct two experiments that aim to demonstrate the framework and offer empirical evidence of the models´ performance. In the first evaluation, we generate a dataset and assess predictive accuracy on the test set and on what-if analysis, providing later a conceptual interpretation of models for the different levels of the framework. In the second, we evaluate model performance on four benchmark datasets for prediction and what-if analysis. The evaluation metrics are Mean Absolute Error **(MAE**), Root Mean Squared Error (**RMSE**), and we use for relative performance the Weighted Absolute Percentage Error (**WAPE**), a metric more robust to divisions by small values than the commonly used Mean Absolute Percentage Error (Hewamalage et al., 2023).

1 The code and datasets in the evaluations are available for complete reproducibility in the repository: https://github.com/drzg15/inherently_interpretable_causal_machine_learning.

# 4 Experiments

## 4.1 Evaluation 1

We present an example of a production organization with different measurable factors that interact with each other. These variables are batch size *(size),* productivity index *(productivity),* material properties *(material)*, energy consumption *(energy),* and number of workers *(personnel).* One observation corresponds to a production batch. The data-generating process is shown on the left side of Figure 5.The equations include linear, sinusoidal, logarithmic components and interactions. The noise terms are drawn from Gaussian distributions ($\mathcal{N}$), where the first parameter denotes the mean and the second parameter denotes the standard deviation. The noise terms differ in their standard deviations to reflect that some variables exhibit greater inherent variability, are measured on different scales, and correspond to plausible magnitudes in a real industrial setting. We generate 8,000 observations for training and the same number for testing.

The right side of Figure 5 depicts the causal graph derived from the equation terms; if one variable appears within the formula of another, it is a parent of it, hence a causal link is drawn in that direction. The goal is to analyze the *energy* consumption of the production system for sustainability decisions.

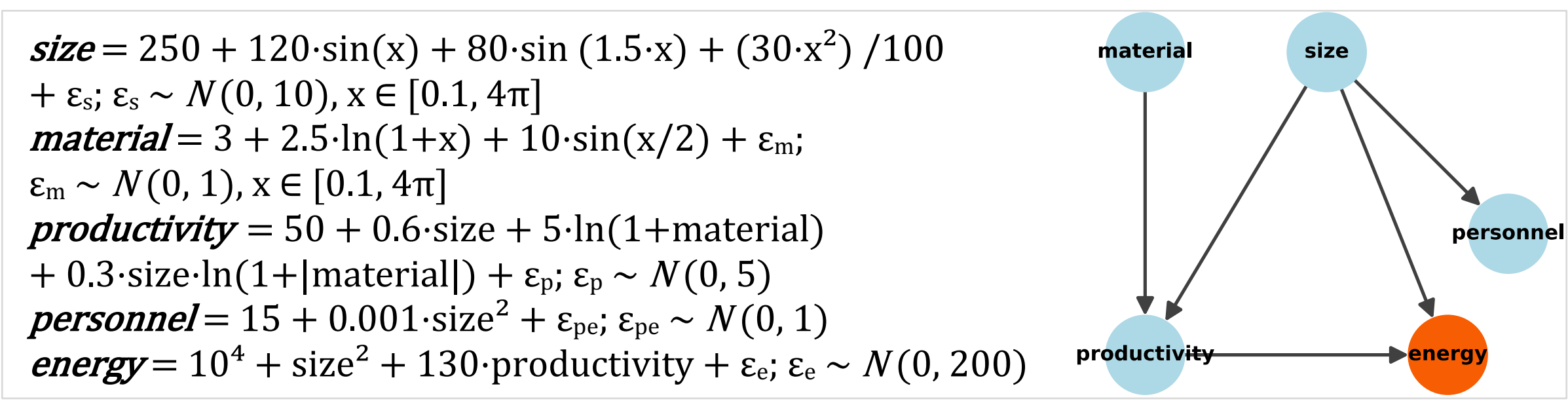

*Figure 5. Data generating process and causal graph for the production use case.*

The ML models are trained with all variables, while the CML models are trained following the causal graph, modelling nodes only with its parents (i.e., *energy, personnel, productivity*). The root nodes are exogenous inputs rather than fitted models. The trained models are first used to predict the test set.

Afterward, we perform a what-if analysis with 8,000 observations by asking: What would the *energy* consumption be if every batch has a size of 300 units? do(size =300), while keeping the other variables as in the training data. The ground truth are the results of the intervention from the data-generating process. The ML models predict the effect of the intervention by using the same inputs as in training, except with the intervened batch *size*, while the CML propagates the effects of intervention by using the causal graph.

The results are shown in the units of the target. Table 1 presents the results for the test set sorted by MAE. The "type" column shows the training approach, non-causal **ML,** and **CML**. In the test set, most of the models achieved a high performance with a WAPE below 13%. Here, none of the model´s types outperformed systematically the others. The CML-SR and GAM were slightly better than their ML counterparts, and the CML-MLP and Linreg performed worst.

Table 2 shows the results of the what-if analysis. All models performed worse than on the standard test-set prediction task, as what-if queries are inherently more challenging. In this setting, all CML models outperformed the purely predictive ML approaches. Notably, even the simplest CML model (CML-Linreg) achieved better performance than the best ML model (RForest). The performance gap between CML and its ML counterpart is substantial, about 8 percentage points in WAPE for the RForest and even more than 41 percentage points for the SR.

| Type | Model | MAE | RMSE | WAPE |
|---|---|---|---|---|
| CML | SR | 161.7618 | 201.7267 | 0.1381 |
| CML | GAM | 172.7371 | 221.7026 | 0.1474 |
| ML | GAM | 258.9799 | 328.4655 | 0.2210 |
| ML | Linreg | 871.7708 | 1,091.9058 | 0.7440 |
| ML | SR | 873.2224 | 1,087.1806 | 0.7452 |
| ML | MLP | 2,350.8373 | 3,917.5160 | 2.0063 |
| ML | XGBoost | 4,014.8151 | 5,945.5700 | 3.4263 |
| CML | XGBoost | 4,159.9224 | 6,194.8302 | 3.5502 |
| CML | RForest | 8,946.8788 | 12,131.4808 | 7.6354 |
| ML | RForest | 9,150.5342 | 12,308.9708 | 7.8092 |
| CML | Linreg | 11,134.4783 | 14,122.7942 | 9.5024 |
| CML | MLP | 14,807.5726 | 26,018.5234 | 12.6371 |

*Table 1. Results for the test set.*

| Type | Model | MAE | RMSE | WAPE |
|---|---|---|---|---|
| CML | SR | 546.3665 | 685.4107 | 0.3378 |
| CML | GAM | 655.6439 | 818.8543 | 0.4054 |
| CML | MLP | 966.1252 | 1,200.3433 | 0.5974 |
| CML | RForest | 1,097.4191 | 1,510.6087 | 0.6786 |
| CML | XGBoost | 1,199.0493 | 1,440.4911 | 0.7414 |
| CML | Linreg | 10,011.9696 | 10,259.0133 | 6.1909 |
| ML | RForest | 14,013.0083 | 15,039.8835 | 8.6649 |
| ML | GAM | 16,522.3962 | 18,800.2062 | 10.2166 |
| ML | XGBoost | 18,591.6119 | 20,493.5576 | 11.4961 |
| ML | Linreg | 67,071.1320 | 75,544.8099 | 41.4733 |
| ML | MLP | 68,350.0023 | 76,170.1547 | 42.2640 |
| ML | SR | 68,489.7697 | 77,121.4824 | 42.3505 |

*Table 2. Results for the what-if analysis.*

Now we interpret the trained models in the different levels of the framework. Figure 6 (a) depicts the first-level, association post-hoc interpretation, by using summary SHAP values of the XGBoost. The results indicate that batch *size* was the most important predictor, followed by *productivity* and *personnel*, all affecting the target in the positive direction.

Figure 6 (b) and (c) illustrate the second level of the framework, inherently interpretable associations. The equation in (b) corresponds to the one found by the SR to describe the data, which shows a simple formula with *productivity* and *personnel*. Figure 6 (c) depicts the partial dependence plots (PDPs) of the GAM, making transparent how the individual predictor values affect the target variable and shape of these relationships. The visuals reveal that *material* and *personnel* have a complex shape to describe *energy*.

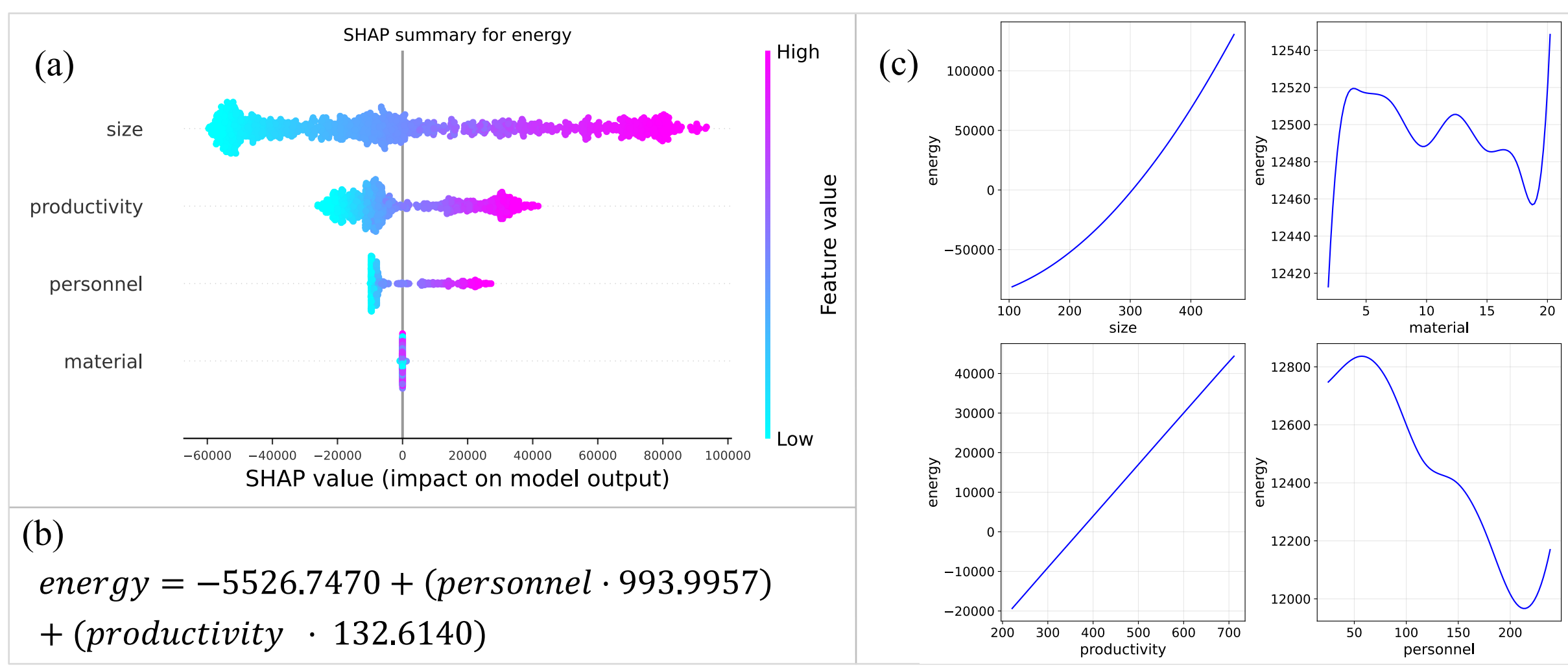

*Figure 6. Post-hoc and inherent interpretation associations. (a) SHAP summary for the XGBoost; (b) equation found by the SR; (c) GAM´s PDPs.*

While these methods provide understanding, and in the case of the IIML models direct transparency regarding how variables influence model decisions, their interpretations can be misleading. All models assign substantial importance to the *personnel* variable, even though it has no causal link to the target outcome. Due to confounding with *size*, the models automatically selected *personnel* based on its strong correlation with the target. Unable to disentangle this spurious association, the models attributed a large predictive effect to this variable (and less effect to the other variables), which yielded strong test-set performance but poor results for the what-if analysis. Such interpretations, although highly transparent with the IIML models, risk distorting decisions.

Next, we present interpretations of the causal models that explicitly leverage the causal graph during training. To show the level 3 of the framework, causal post-hoc interpretation, Figure 7 illustrates SHAP values for the XGBoost models trained in accordance with the causal graph. We display a visual for each model for the variables with parents, namely *personnel, productivity and energy*.

In this case, the variable of *personnel* does not appear in the explanation of *energy* since it has no causal link to the target. Importantly, in contrast to Figure 6 (a), the model for *energy* concentrated prediction power on the two causal variables because it cannot be misled by the spurious correlation with *personnel*.

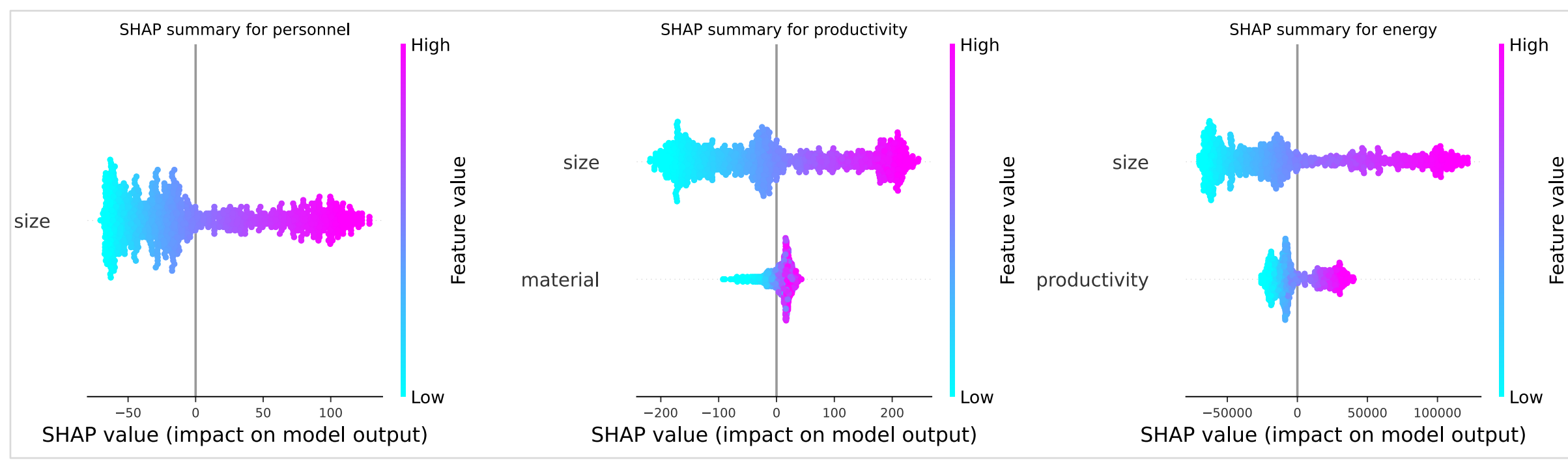

*Figure 7. SHAP summary graph for the XGBoost models trained based on the causal graph.*

Now we present the interpretation at the top level of the framework, the causal inherent interpretation. Figure 8 (a) shows the causal graph with the equations discovered through the symbolic regression (SR). Since the SR concentrated only the causal predictors, it could recover nearly all functional terms with different coefficients from the data-generating process, except one related to *productivity*. This is in stark contrast to Figure 6 (b), where the model did not match any functional term from the data-generating process and some causes where missing.

Figure 8 (b) shows the PDPs of the GAMs, one model in each row. These plots reveal that the functional forms of the GAMs diverge from the ML approach, reflecting adjustments imposed by the causal structure. Because *personnel* and *material* are not predictors of *energy*, their complex associations with the target shown in Figure 6 (c) are absent. As a result, the models produced by combining GAMs with the causal graph are simpler to understand.

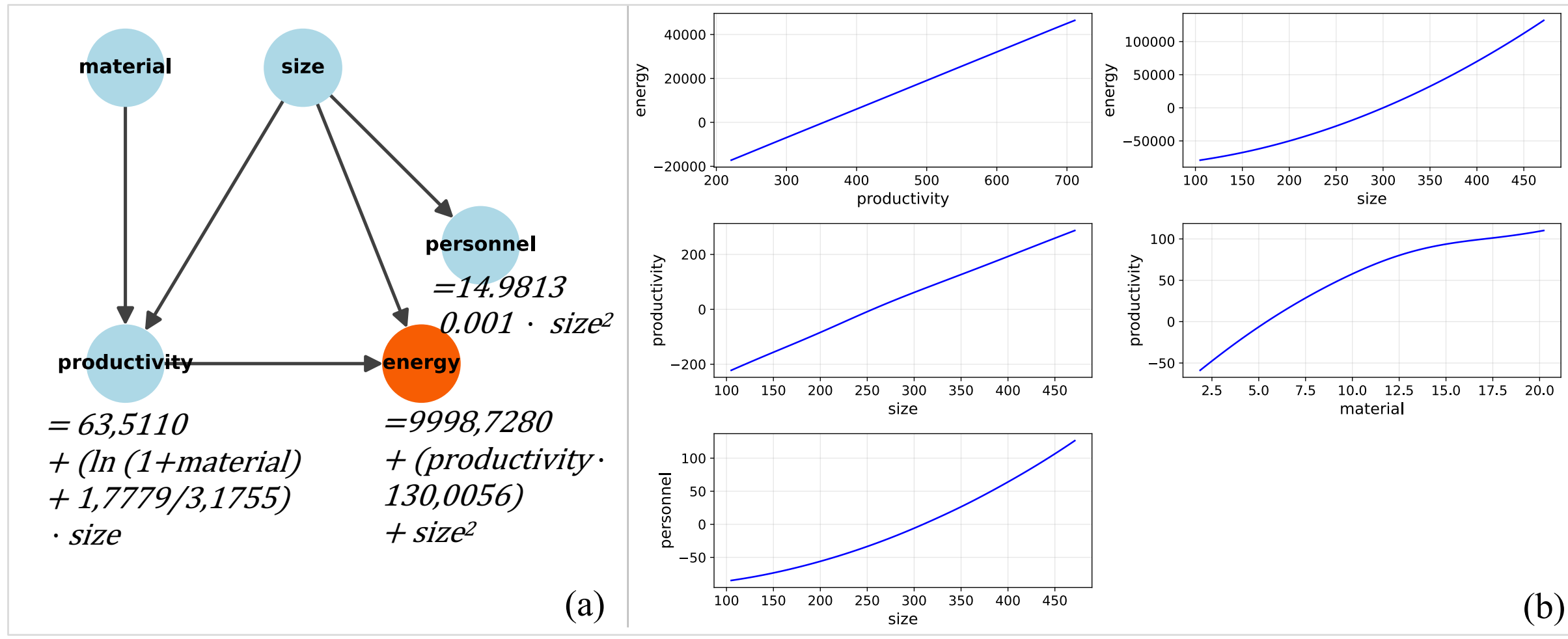

*Figure 8. Causal inherent interpretation. (a) Causal graph with SR equations; (b) GAM PDPs.*

## 4.2 Evaluation 2

In this subsection, we aim to obtain more empirical evidence on the performance of CML models compared to standard ML models for prediction and what-if analysis. Also, we further evaluate the performance of the IIML approach relative to the black-box ML methods. To this end, we use datasets from Geffner et al. (2022), that aim to benchmark CML models. Each dataset includes a ground truth causal graph, 4,000 training observations, 2,000 test set samples, and the same for intervention sets.

We consider four datasets with continuous targets: The *nonlin_simpson* and *symprod_simpson* datasets each contain 4 variables connected by 4 causal edges. The *large_backdoor* dataset has 9 variables and 10 edges, while the *weak_arrows* includes 9 variables and 15 edges. All datasets feature intricate, nonlinear causal relationships. The causal graphs and equations can be found in the code repository.

The interventions in the datasets are used to evaluate the what-if scenarios. The results for each dataset are presented individually in the repository. Here, we report the aggregated results obtained by normalizing the predictions and true values and computing the median of the metrics across datasets.

The test set results are presented in Table 3, showing that the ML-GAM model achieved the best performance, followed by MLP and XGboost. The difference between the ML and CML variants was minimal, typically under one percentage point in WAPE for nearly every algorithm. Table 4 reports the results for the what-if analysis; here, all the models had a worse performance that in the test set. In this case, the CML-MLP was the best model. Also, all CML models outperformed their ML counterparts for more than 6 percentage points in WAPE.

| Type | Model | MAE | RMSE | WAPE |
|---|---|---|---|---|
| ML | GAM | 0.1898 | 0.2374 | 24.3297 |
| ML | MLP | 0.1900 | 0.2387 | 24.3636 |
| ML | XGBoost | 0.1907 | 0.2400 | 24.4475 |
| ML | RForest | 0.1928 | 0.2424 | 24.7198 |
| CML | GAM | 0.1950 | 0.2434 | 24.9916 |
| CML | MLP | 0.1957 | 0.2445 | 25.0801 |
| CML | RForest | 0.1972 | 0.2462 | 25.2799 |
| CML | XGBoost | 0.1980 | 0.2475 | 25.3799 |
| CML | SR | 0.2109 | 0.2671 | 27.0674 |
| ML | SR | 0.2145 | 0.2722 | 27.5077 |
| ML | Linreg | 0.2510 | 0.3152 | 29.7925 |
| CML | Linreg | 0.2842 | 0.3643 | 35.1300 |

*Table 3. Exp 2. Results for the test set.*

| Type | Model | MAE | RMSE | WAPE |
|---|---|---|---|---|
| CML | MLP | 1.0406 | 1.3637 | 104.0650 |
| CML | XGBoost | 1.0407 | 1.3721 | 104.0680 |
| CML | GAM | 1.0465 | 1.3690 | 104.6519 |
| CML | RForest | 1.0482 | 1.3870 | 104.8175 |
| CML | Linreg | 1.0850 | 1.3180 | 108.4992 |
| CML | SR | 1.0853 | 1.3526 | 108.5327 |
| ML | XGBoost | 1.1224 | 1.4292 | 112.2403 |
| ML | GAM | 1.1231 | 1.4337 | 112.3113 |
| ML | MLP | 1.1275 | 1.4411 | 112.7521 |
| ML | RForest | 1.1403 | 1.4528 | 114.0344 |
| ML | SR | 1.1476 | 1.4571 | 114.7591 |
| ML | Linreg | 1.1629 | 1.4464 | 116.2874 |

*Table 4. Exp. 2. Results for the interventions.*

# 5 Discussion

Motivated by the growing use of machine learning for decision support (Abdel-Karim et al., 2021; Jayatilake & Ganegoda, 2021; Jhaveri et al., 2022), the need to move beyond prediction toward what-if analysis (Hünermund et al., 2022), and the advantages of inherent model interpretability rather than post-hoc explanations (Rudin, 2019; Zschech et al., 2025), we proposed the combination of principles from causal inference with inherently interpretable models.

Addressing **RQ1**, we proposed a framework that integrates Structural Causal Models (SCMs) with inherently interpretable machine learning (IIML) to create causal and interpretable predictive models for cross-sectional data. For **RQ2**, we compared the performance of the causal IIML approach with ML models trained with all available data in two experiments: one to demonstrate the levels of interpretability and the second to show the performance of the models with benchmark datasets.

In the proposed framework and demonstration, we distinguished between several levels of interpretability: association post-hoc interpretation, association inherent interpretation, causal post-hoc interpretation, and causal inherent interpretation. These levels illustrate a gradual progression toward deeper interpretability in ML modelling. The last level provides a clear understanding of the variables involved, explicit causal representation of the relationships among them, and an interpretable functional form describing how these variables interact.

The causal graph avoids biases and spurious correlations, ensuring not only robust predictive performance but enabling the second level of the causality ladder, doing or intervention (Pearl, 2018). Our results show that across all evaluations, the causal models achieved competitive predictive performance on the test set while also enabling reliable what-if analysis. In contrast, the machine learning models performed considerably worse in the what-if evaluations. These findings suggest that incorporating causal structure into modeling is essential for tasks that extend beyond pure prediction.

In this regard, modelling according to the causal graph improved all models' ability to capture the functional relationships among variables, concentrating predictive power on the relevant causal factors. Also, in the case of the SR, the results followed closer approximation to the underlying data-generating process, and for the GAM, the resulting functions had simpler forms. This suggests that incorporating causal structure not only yields causal insights but also makes individual models easier to interpret and more closely aligned with the data-generating process.

Furthermore, the IIML models (GAMs and SR) also had a competitive performance compared to their complex black-box counterparts, being even the best models in several tasks. This underlines the idea that it is possible to achieve high predictive accuracy for tabular data comparable to more complex models without sacrificing interpretability (Bohlen et al., 2025; Rudin, 2019; Zschech et al., 2025).

Regarding the IIML models, both SR and GAMs offer strong interpretability through different approaches and have a competitive performance in the evaluation tasks. GAMs provide visual insights into the relationships between variables using partial dependence plots, while SR presents the explicit mathematical formulas that describe these relationships. We do not claim that one approach is superior to the other; rather, they may appeal to different target audiences depending on their analytical needs and preferences.

Although causal models can be simpler and more interpretable, since they typically involve fewer variables, training them can be more expensive overall. This is because one model must be trained for each node (conditioned on its parents in the causal graph), increasing the total number of models. The cost becomes especially relevant when the causal graph has many variables, and the dataset is large. This it is more pronounced for models that are expensive to train, such as SR (Reinbold et al., 2021; Zschech et al., 2025).

These results contribute to the CML literature by incorporating the use of IIML for modelling causal graphs. Also, we expand ML interpretability literature by introducing a causal framework that enables what-if analysis and allows the models to yield causal insights. For practical applications in decision-making, we recommend first to describe causal graphs to make explicit the assumptions about the data and the causal structure of the system. Later, the modelling can be done with IIML. This choice depends on the desired type of interpretability (e.g., a GAM-based approach vs. SR), as well as the complexity of the causal graph and the amount of available data. This integration ensures maximum interpretability and enhances transparency in the decisions derived from such models.

## 5.1 Limitations and future work

The study has several limitations that can be addressed in future research. First, our work primarily addressed the first (seeing) and second rung of the Ladder of Causation (doing), but future research could aim to reach the third rung (counterfactual reasoning), which requires the structural causal model to be invertible with respect to the noise component. Future research could address this, but also incorporate the modelling of the noise component with IIML models.

Also, an important limitation is the use of the correct causal graphs and do not empirically examine what happens when the causal graphs are misspecified. In the evaluations, we rely on ground truth causal graphs, and this benefits the causal models by making them simpler and reducing the modelling complexity. However, in real-world applications, the exact causal graph is typically unknown and must be constructed based on domain knowledge or inferred from data. As discussed in the literature, tools exist to support the extraction and reliability assessment of causal graphs (Cinelli & Hazlett, 2020; Eulig et al., 2025; Faller et al., 2024). These causal graphs may not always be accurate and cannot be fully verified in practice. In this regard, misspecification of the graph may affect the quality of the resulting models, accuracy of the predictions, and specially the evaluation of what-if scenarios. Future work could therefore investigate the sensitivity of the proposed approach to graph misspecification, assess its robustness under graph-structure uncertainty, and how this can affect interpretability and reliability in decision-making.

Furthermore, we discussed the interpretability conceptually in the demonstration and did not qualitatively evaluate it through expert assessment; even when models are inherently transparent, some users may still find them difficult to interpret in practice. For instance, SR equations might become overly convoluted with many variables and intricate operations, and GAM´s shape plots can become complex when there are interaction terms or highly nonlinear effects. Future empirical research could systematically assess interpretability and human trust by examining how different stakeholders perceive, understand, and apply inherently interpretable causal models. In particular, such studies could investigate the cognitive demands associated with analyzing GAM shape plots or SR equations, and how this complexity affects their effective use in decision support settings.

Additionally, this study focused on what-if analysis for decision support rather than on the estimation of causal effects, which requires dedicated methods from causal machine learning (Feuerriegel et al., 2024), such as double machine learning (Chernozhukov et al., 2018) and meta-learners (Künzel et al., 2019). Future research should examine how such approaches can be integrated with IIML models.

Also, in this study, we did not conduct a detailed analysis of why the different models performed as they did, for example, why the IIML models achieved such strong results, while traditional black-box models for tabular data, such as tree-based Random Forests and XGBoost, were not the best performers. Future research could investigate these performance differences in relation to the characteristics of the underlying data-generating process.

Finally, the proposed framework in this paper focuses on cross-sectional data. Future research could extend the framework to time-series settings and investigate how temporal dependencies, lag structures, and dynamic confounding can be appropriately incorporated, modelled, and interpreted with IIML models.

# 6 Conclusions

In this paper, we highlighted the advantages of combining causal inference with inherently interpretable machine learning, proposed a framework for their integration for cross-sectional data, and evaluated the approach with several experiments. We found that models trained with causal graphs achieved a competitive performance for test set evaluation and excel at what-if scenario evaluation. Also, the inherently interpretable models trained within a causal structure achieved competitive predictive performance compared to complex black-box models, while providing a higher level of transparency and interpretability.

This study is a first step to bridge CML and IIML, two fields that naturally complement each other to offer interpretability and causality. We hope that this integration will encourage further research to foster more transparent, trustworthy, and reliable ML-driven decision support.

# 7 Acknowledgements

This study was supported by the Federal Ministry for Economic Affairs and Energy of Germany under Grant No. 03EN21094 (HeatTransPlan), and we gratefully acknowledge their support.